\documentclass[%
reprint,
superscriptaddress,
nobibnotes,
amsmath,amssymb,
aps,
prb,
]{revtex4-1}

\usepackage{graphicx}
\usepackage{dcolumn}
\usepackage{bm}
\usepackage{hyperref}
\hypersetup{
    breaklinks=true,
    bookmarks=true,
    pdfauthor={},
    pdftitle={},
    colorlinks=true,
    citecolor=blue,
    urlcolor=blue,
    linkcolor=magenta,
    pdfborder={0 0 0}
}

\usepackage{mathtools} 
\usepackage{upgreek} 

\usepackage{color,soul} 
\usepackage[normalem]{ulem} 
\usepackage{xfrac} 
\usepackage{braket} 
\usepackage{cases} 
\usepackage{placeins} 
\usepackage{booktabs} 

\usepackage{floatrow} 
\usepackage[caption=false]{subfig}  

\newcommand{\rx}{r_\mathrm{X}}
\newcommand{\ry}{r_\mathrm{Y}}
\newcommand{\rSb}{r_\mathrm{Sb}}
\newcommand{\rM}{r_\mathrm{M}}

\begin{document}

\floatsetup[figure]{style=plain,subcapbesideposition=top} 
\floatsetup[table]{style=plain, capposition=top} 

\preprint{APS/123-QED}

\title{\textit{Ab initio} study of the crystal and electronic structure of mono- and bi-alkali antimonides: Stability, Goldschmidt-like tolerance factors, and optical properties}


\author{J. K. Nangoi}
\email[Corresponding author: jn459@cornell.edu]{}
\affiliation{
    Department of Physics, Cornell University, Ithaca, New York 14853, USA
}

\author{M. Gaowei}
\affiliation{
    Brookhaven National Laboratory, Upton, New York 11973, USA
}

\author{A. Galdi}
\affiliation{
    DIIn Department of Industrial Engineering, Universit\`{a} degli Studi di Salerno, Fisciano (SA), Italy
}

\author{J. M. Maxson}
\affiliation{
    Department of Physics, Cornell University, Ithaca, New York 14853, USA
}

\author{S. Karkare}
\affiliation{
    Department of Physics, Arizona State University, Tempe, Arizona 85287, USA
}

\author{J. Smedley}
\affiliation{
    SLAC National Accelerator Laboratory, 2575 Sand Hill Rd, Menlo Park, CA 94025
}

\author{T. A. Arias}
\affiliation{
    Department of Physics, Cornell University, Ithaca, New York 14853, USA
}


%


\begin{abstract}

    Mono- and bi-alkali antimonides, X$_2$YSb (X and Y from Group I), are
    promising for next-generation electron emitters due to their capability of
    producing high-quality electron beams. However, these materials are not yet
    well understood,
    in part due to the technical challenges in growing pure, ordered alkali
    antimonides. For example, in the current literature there is a lack of
    complete understanding of the mechanically stable 
    crystal structures of these materials.
    As a first step towards understanding this issue, this paper presents an
    \textit{ab initio} study of stability of single-crystal mono- and bi-alkali
    antimonides
    in the $D0_3$ structure, the
    structure generally assumed in the literature for these materials.  Finding
    that many of these materials actually are unstable in the $D0_3$ structure, we
    formulate a new set of Goldschmidt-like tolerance factors
    that accurately predict $D0_3$ stability using a procedure analogous to
    machine-learning perceptron-based analysis. Next, we consider possible stable  structures for materials that we predict to be unstable in the $D0_3$ structure. Taking as examples the mono- and bi-alkali antimonides Cs$_3$Sb and Cs$_2$KSb, which also are technologically 
    interesting for photoemission and photoabsorption applications,
    respectively, we note that the most unstable phonon displacements are consistent with 
    the cubic structure,
    and we therefore perform extensive {\em ab initio} searches to identify potential ground-state structures in a cubic lattice. 
    Our X-ray
    diffraction experiments confirm that indeed these two materials are not stable in the $D0_3$ structure and show scattering that is 
consistent with our new, proposed stable structures. Finally, we explore \emph{ab initio} the implications of the breaking of the $D0_3$ symmetry on the electronic structure, showing significant impact on the location of the optical absorption edge.

\end{abstract}


\maketitle



\section{Introduction} \label{sec:intro}

Alkali antimonides have been, and continue to be of great interest as potential
high-quality electron emitters for various applications, including electron
accelerators and ultrafast electron diffraction and
microscopy.\cite{ref:zunger_m3sb, ref:schubert_accel, ref:cultrera_cold-cs3sb,
ref:schubert_xrd, ref:epitaxial-cs3sb, ref:whl-ued-NaKSb}
Despite their promise, these materials are not very well-understood, in part
because growth of single-crystal alkali antimonides remains very challenging:
the resulting materials are often polycrystalline or disordered,\cite{ref:schubert_accel, ref:epitaxial-cs3sb} thereby
limiting the understanding of the equilibrium crystal structures and thus of
the ultimate promise of these materials as electron
emitters. 
Moreover, 
even once successfully grown, single-crystal versions of these
materials are extremely sensitive to vacuum conditions and
can survive only in ultra-high-vacuum in their thin film forms, making
experimental studies of their structural and optoelectronic properties extremely
challenging.\cite{ref:epitaxial-cs3sb}

\begin{figure}[h!]
    \centering
    (a) \hspace{10em} (b)
    \\
    \includegraphics[width=0.43\linewidth]{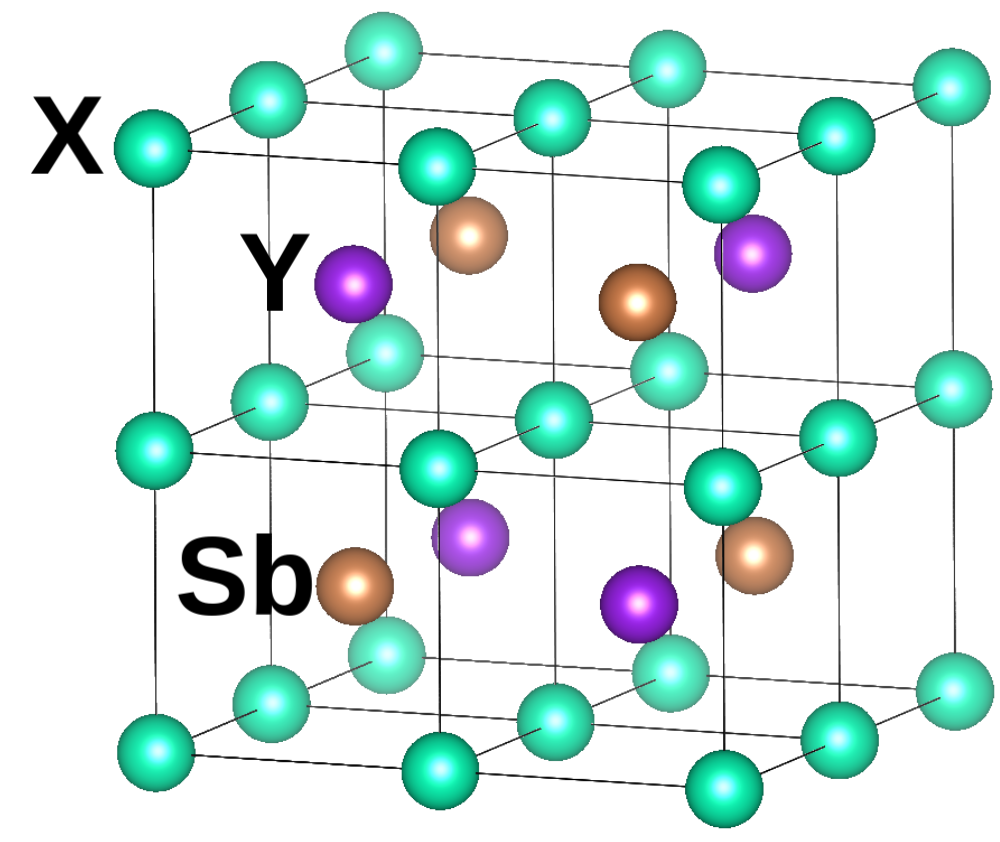}
    \includegraphics[width=0.43\linewidth]{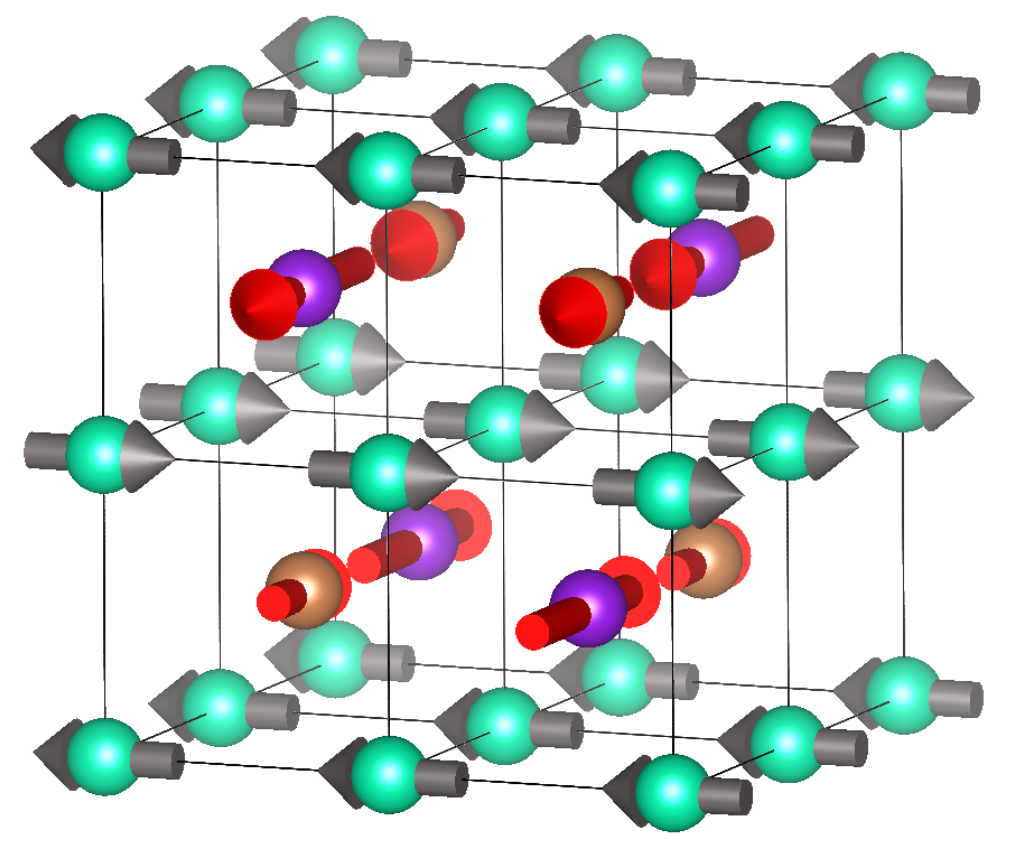}
    \\
    (c) \hspace{10em} (d)
    \\
    \includegraphics[width=0.43\linewidth]{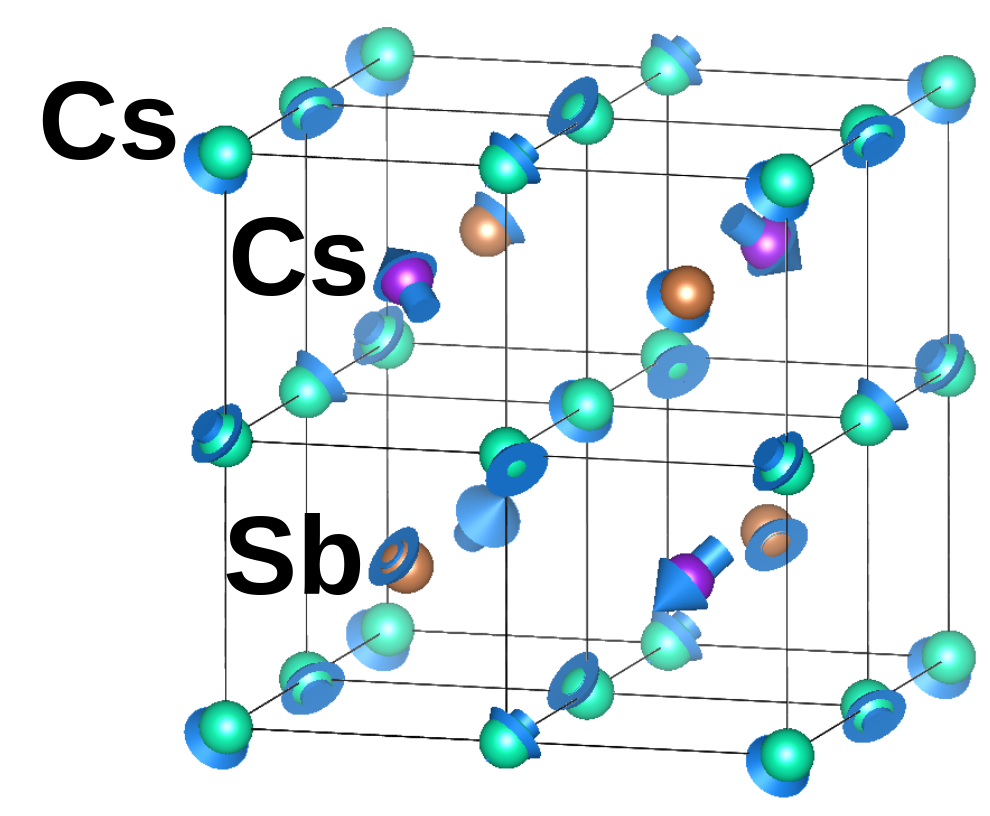}
    \includegraphics[width=0.43\linewidth]{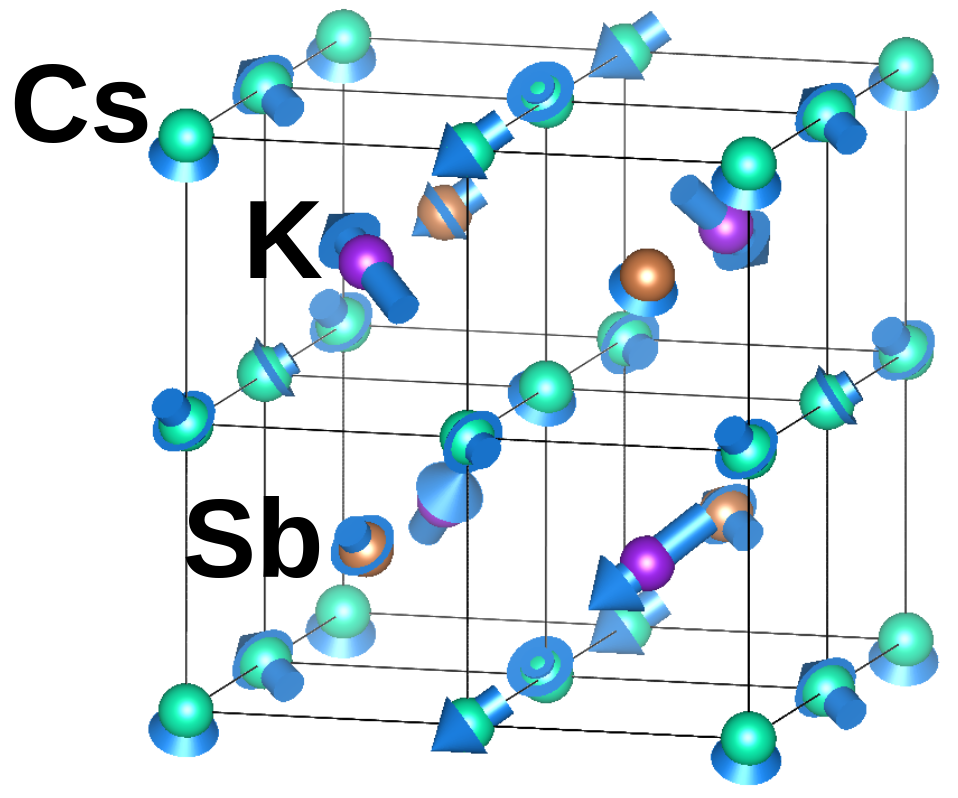}

    \caption{
        (Color online) 
        (a) $D0_3$ structure of mono- and bi-alkali antimonides (X$_2$YSb), 
        (b) complex phonon polarization vector of an imaginary branch at the \emph{\textsf{X}}-point in the face-centered-cubic Brillouin zone (corresponding to phonon wavevector directed upwards) found in all materials under study that exhibit imaginary phonon frequencies,
        (c) atomic displacements from the $D0_3$ structure to the stable structure of Cs$_3$Sb (X = Y = Cs),
        and (d) atomic displacements from the $D0_3$ structure to the stable structure of Cs$_2$KSb (X = Cs, Y = K): 
        Group I elements (X, green spheres, and Y, purple spheres), 
        antimony (Sb, brown spheres),
        real part of phonon polarization vector (gray arrows), imaginary part of phonon polarization vector (red arrows), atomic displacements 
        (blue arrows).
    }

    \label{fig:xtal-structs}
\end{figure}

Previous work has generally assumed the $D0_3$ structure 
(Fig.~\ref{fig:xtal-structs}a)
as the crystal structure for mono- and bi-alkali antimonides, X$_2$YSb (X and Y
are Group I alkali metals).\cite{ref:orig-D03, ref:christensen-1985,
ref:zunger_m3sb, ref:ettema-2000, ref:ettema-2002, ref:kalarasse_optical,
ref:murtaza_bi-alk-ant, ref:finkenstadt_cs3sb, ref:cocchi-2019-condens,
ref:CsKSb-GW-gaps, ref:cocchi-2021-elec, ref:cocchi-2021-micro} 
However, a very recent publication on
the first ever successful
epitaxial growth of Cs$_3$Sb finds that,
although the thin-film structure is cubic and single crystal, it is not clear
whether the structure is $D0_3$ or some other cubic phase with lower
symmetry.\cite{ref:epitaxial-cs3sb}
To complicate the matter, the Materials Project database\cite{ref:matProj} reports imaginary phonon frequencies and thus mechanical instability for Cs$_3$Sb
in the $D0_3$ structure.

To clarify the crystal structures of all possible mono- and bi-alkali antimonides
(X$_2$YSb), we report below the results of a detailed \textit{ab initio} study
of the mechanical stability of twenty-five X$_2$YSb materials in the $D0_3$
structure.  For X and Y, we explore Li, Na, K, Rb, and Cs, the five alkali
metals most likely to be used as electron
emitters.\cite{ref:spicer_rb3sb-etc,ref:gaoxue_qe-lifetime} (We exclude Fr
because it is impractical, as its most stable isotope has a very short
half-life of merely 21 minutes.\cite{ref:francium-half-life})
Our predictions of the stability of these materials in the $D0_3$ structure show a clear
pattern corresponding to multiple machine-learning perceptron
criteria\cite{ref:perceptron} that lead us to propose a set of physically motivated
Goldschmidt-like tolerance factors\cite{ref:goldschmidt} for the stability of
the X$_2$YSb materials.

Next, to better understand potentially stable crystal
structures, we choose to further study two of our predicted $D0_3$-unstable materials, Cs$_3$Sb and
Cs$_2$KSb, 
which are technologically interesting for photoemission\cite{ref:cultrera_cssb-emittance} and photoabsorption\cite{ref:ettema-2002, ref:kalarasse_optical} applications, respectively.
For these materials, 
we present the results of an extensive search for stable cubic structures, finding full consistency with our X-ray diffraction measurements.
Finally, we present \emph{ab initio} results on the impact of the resulting breaking of $D0_3$ symmetry in these materials on their electronic structure and optical absorption properties.


\section{Procedure} \label{sec:procedure}

Here, we describe our overall \emph{ab initio} procedures, leaving additional details
in the Supplemental Material.\cite{ref:supplemental}
%
\nocite{ref:pbe}
\nocite{ref:oncv}
\nocite{ref:sg15}
\nocite{ref:wannier}
%
The \textit{ab initio} study of mechanical stability performed in this work
requires electronic structure calculations and phonon dispersion calculations
of the X$_2$YSb materials.  All electronic structure calculations employ the
plane-wave density-functional theory framework as implemented in the
JDFTx software package.\cite{ref:jdftx} All phonon calculations employ a modified
version of the frozen phonon method, which allows calculations of phonons at
arbitrary wave vectors using a real-space method, as implemented within
JDFTx.\cite{ref:jdftx}

To determine the mechanical stability of X$_2$YSb materials in the $D0_3$
structure, we employ the following procedure. First, we relax the atoms and the
lattice from the $D0_3$ structure as a starting point, using 
face-centered cubic (fcc) primitive cell. 
We find 5 materials to be unstable with respect to
these relaxations. Next, to identify instability at wavelengths incommensurate
with
our primitive cell, for the remaining materials we calculate the phonon
dispersion relations to identify any imaginary frequencies, finding an
additional 10 unstable materials. After tabulating the results, we find that
stability of the X$_2$YSb materials follows criteria describable through
machine-learning perceptron-based analysis.\cite{ref:perceptron} Due to the
small data set of 25 materials, we do not require the perceptron algorithm
itself and determine the criteria manually.

Next, to identify stable structures for the materials that are unstable in the
$D0_3$ structure, we focus on Cs$_3$Sb and Cs$_2$KSb, searching for stable
cubic structures through multiple sets of perturbation-relaxations as follows.
For each set, we begin with the atoms in the $D0_3$
structure in the conventional cubic unit cell, then we
uniformly displace the atomic locations by 0.05~$\mathrm{\AA}$, and
subsequently relax the atoms until all forces are $\leq$ $\sim$0.01~eV/\AA. 
We repeat this procedure until
we identify a mechanically stable structure as confirmed by calculation of the resulting phonon dispersion relations.

For comparison with our X-ray measurements, we then
calculate the powder X-ray diffraction patterns of the above resulting stable structures using
VESTA.\cite{ref:vesta}
The details of the experimental setup and the growth and X-ray measurement
conditions of our samples are given elsewhere.\cite{ref:chess-xrd,
ref:Cs-K-Sb-growth}

Finally, to study optical absorption properties, we compute linear optical absorption
coefficients by first calculating the contributions of direct photoexcitation
processes to the imaginary part of the dielectric constant, $\epsilon_2$, by
employing the Wannier interpolation method\cite{ref:wannier_review} and
Monte Carlo integration as described in Ref.~\onlinecite{ref:shankar_acs}. We
then use the resulting $\epsilon_2$ as a function of photon frequency $\omega$
to calculate the linear absorption coefficient $\alpha(\omega)$ as described in
Ref.~\onlinecite{ref:optical-properties-calc}. 


\section{Results and Discussion} \label{sec:results}


\subsection{Stability of the $D0_3$ structure} \label{subsec:results-stab}

Relaxing within a primitive face-centered cubic (fcc) cell immediately reveals the following five materials (out of the twenty-five under consideration) to be unstable in the $D0_3$ structure: K$_2$LiSb, Rb$_2$LiSb, Rb$_2$NaSb, Cs$_2$LiSb, and Cs$_2$NaSb.
The phonon dispersion relations of the remaining materials 
(plots in Supplemental Material\cite{ref:supplemental}) 
reveal ten of those materials to have imaginary phonon frequencies and thus also be mechanically unstable. We further note that all of these materials with imaginary phonon frequencies show instability at the \emph{\textsf{X}}-point in the fcc Brillouin zone.  This then leaves ten materials in our study which are mechanically stable in the $D0_3$ structure.   

Regarding the unstable phonon displacements, we find that the imaginary branches at the \emph{\textsf{X}}-point in the fcc Brillouin zone always exhibit a double degeneracy. The complex phonon polarization vectors of these two modes correspond to either the polarization vector shown in 
Fig.~\ref{fig:xtal-structs}b or to
a second vector generated via 
a 90-degree rotation with respect to the axis along the phonon wave-vector direction.
The real (gray arrows, Fig.~\ref{fig:xtal-structs}b) and imaginary (red arrows, Fig.~\ref{fig:xtal-structs}b) parts of these complex polarization vectors correspond to two distinct unstable phonon displacements of the atoms in real space.
To understand the significance of these displacements, we note that, as shown in 
Fig.~\ref{fig:xtal-structs}a, 
the $D0_3$ crystal structure for X$_2$YSb consists of a simple cubic lattice of X atoms, with all cube body centers occupied by alternating Y and Sb atoms. 
The unstable displacements at the \emph{\textsf{X}}-point in the fcc Brillouin zone thus consist of either shearing of the X atoms relative to the Y and Sb atoms, or shearing of the Y and Sb atoms relative to the X atoms. 
These displacements suggest that the relative size of the X atoms to the Y and Sb atoms plays an important role in determining the stability of the material in the $D0_3$ structure, with the simple-cubic ``cage'' formed by the X atoms being destabilized by the Y and Sb atoms at the cage centers if the Y or Sb atoms are either too large or too small compared to the natural size of the cage.
With this in mind, we summarize our stability results in Table~\ref{tab:stab}, arranging the X and Y atoms in order of their corresponding atomic radii. Confirming our simple hypothesis, indeed the table demonstrates that combinations with either very small or very large Y atoms and with either very small or very large X atoms can be unstable. The next section below provides a more quantitative analysis of these observations.

\begin{table}[h!]

\caption{
    Mechanical stability of X$_2$YSb in the $D0_3$ structure: stable ($\circ$), 
    unstable ($\times$).
    \label{tab:stab}
}

\begin{ruledtabular}

\begin{tabular}{rccccc}
    Y$_1$   & \multicolumn{5}{c}{X$_2$} \\
           & Li$_2$   & Na$_2$   & K$_2$    & Rb$_2$   & Cs$_2$ \\ 
    Cs$_1$ & $\times$ & $\circ$  & $\circ$  & $\times$ & $\times$ \\
    Rb$_1$ & $\times$ & $\circ$  & $\circ$  & $\times$ & $\times$ \\
    K$_1$  & $\circ$  & $\circ$  & $\circ$  & $\times$ & $\times$ \\
    Na$_1$ & $\circ$  & $\circ$  & $\times$ & $\times$ & $\times$ \\
    Li$_1$ & $\circ$  & $\times$ & $\times$ & $\times$ & $\times$
\end{tabular}

\end{ruledtabular}
\end{table}


\subsection{Tolerance factors} \label{subsec:results-tol}

\begin{figure}[h!]
    \centering
    \includegraphics[width=0.9\linewidth]{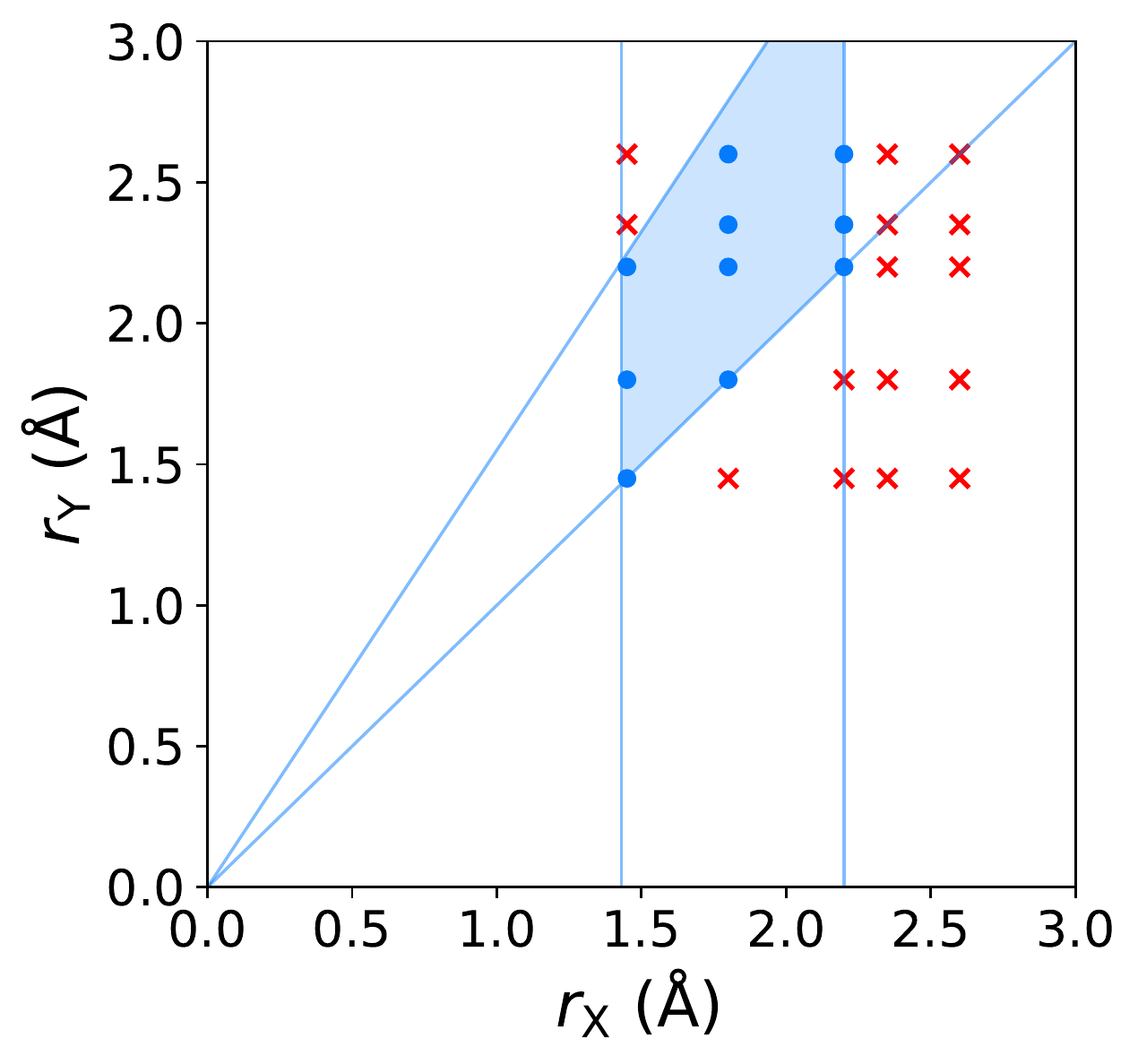}

    \caption{
        Stability versus atomic radii ($\rx$, $\ry$) for all X$_2$YSb materials included in Table~\ref{tab:stab}: materials stable (circles) and unstable ($\times$) in the $D0_3$ structure, region of the stable phase
        (shaded polygon), bounds of the stable-phase region 
        from $1.00 < \ry$/$\rx < 1.55$ (non-vertical lines) and from 
        $0.66 < \rSb $/$\rx < 1.01$ (vertical lines).
    }

    \label{fig:rY-vs-rX}
\end{figure}

To begin the quantitative analysis, Fig.~\ref{fig:rY-vs-rX} displays the results of Table~\ref{tab:stab} in the phase space of the atomic radii\cite{ref:slater_radii} of X ($\rx$) and Y ($\ry$).
As discussed above, we expect there to be upper and lower bounds of stability in terms of the relative size of the X atom to the Y and Sb atoms, corresponding to the ratios $\ry$/$\rx$ and $\rSb$/$\rx$, respectively. 
Therefore, in Fig.~\ref{fig:rY-vs-rX} we consider four perceptron lines, two corresponding to upper and lower bounds for the ratio $\ry$/$\rx$ ($1.00 < \ry$/$\rx < 1.55$) and two corresponding to upper and lower bounds for the ratio $\rSb $/$\rx$ ($0.66 < \rSb $/$\rx < 1.01$).
We see that these four lines indeed separate the stable region from the unstable regions of the phase space, confirming that the relative size of the X atom to the Y and Sb atoms plays a key role in determining the stability of X$_2$YSb in the $D0_3$ structure. 

\begin{figure}[h!]
    \centering
    (a) \\ \includegraphics[width=.939\linewidth]{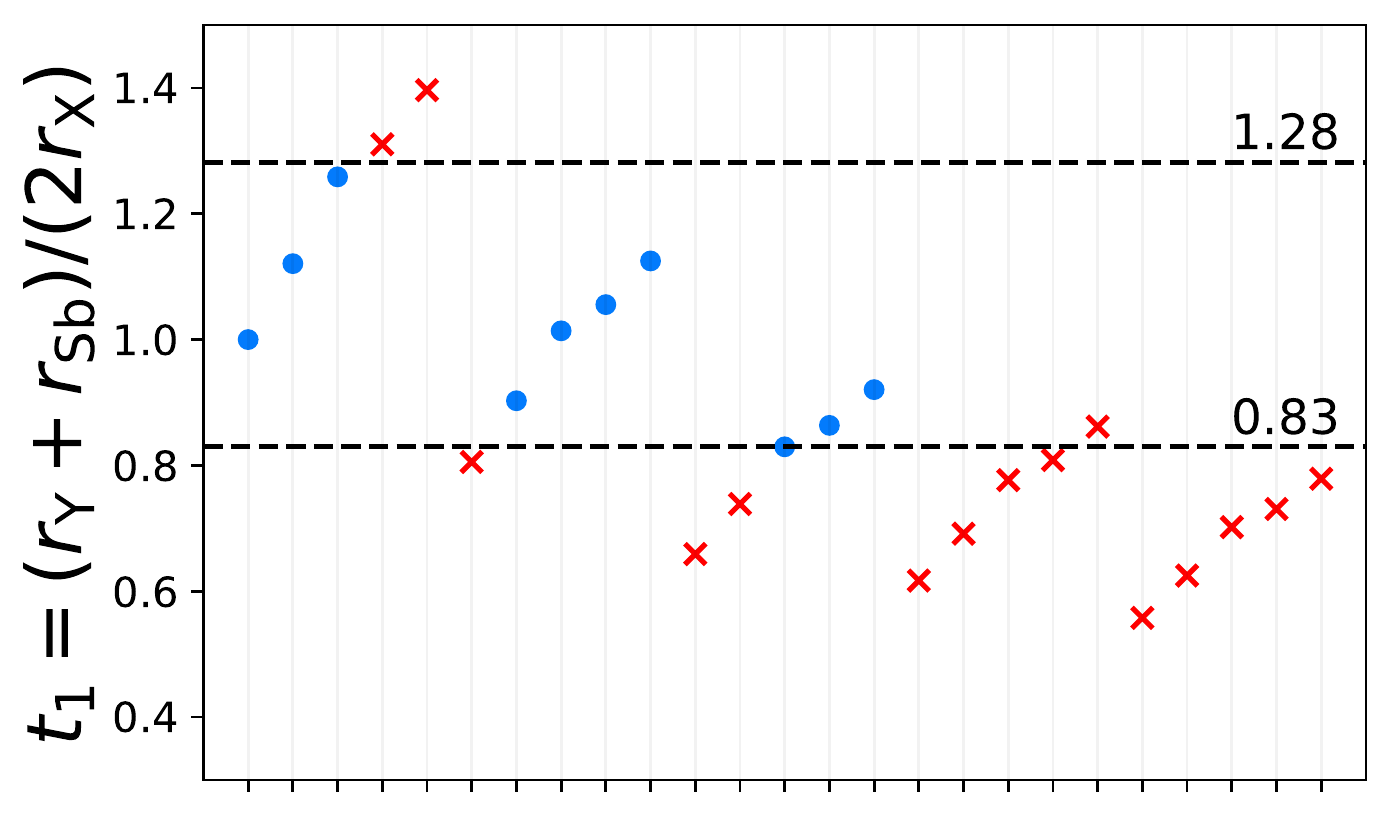}
    \\
    (b) \\ \includegraphics[width=.939\linewidth]{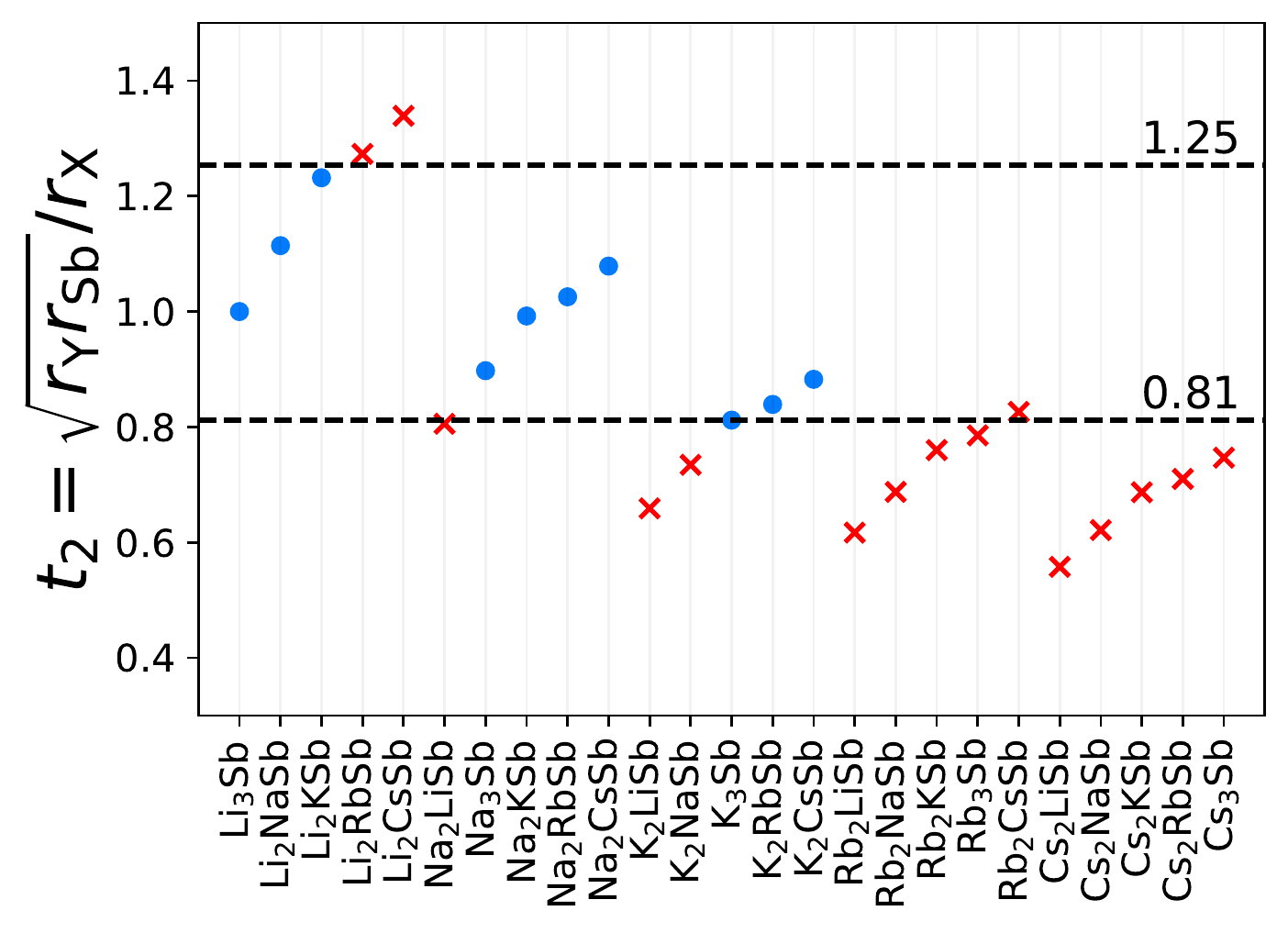}

    \caption{
        Proposed Goldschmidt-like tolerance factors for the X$_2$YSb materials in the $D0_3$ structure:
        materials stable (circles) and unstable ($\times$) in the $D0_3$ structure, 
        bounds of the stable-phase region (dashed lines).
    }

    \label{fig:tol-facs}
\end{figure}

Next, to simplify our results yet further and make them more analogous to the Goldschmidt tolerance factor for perovskites,\cite{ref:goldschmidt} we consider the possibility of constructing one-dimensional tolerance factors.
As discussed above, both ratios, $\ry$/$\rx$ and $r_\mathrm{Sb}$/$\rx$, of the stable materials are bounded.  We therefore
propose the following two simple tolerance factors for determination of stability: the arithmetic mean of the ratios, 
$t_1 \equiv (r_\mathrm{Y} + r_\mathrm{Sb})/(2r_\mathrm{X})$, and the geometric mean of the ratios, $t_2 \equiv \sqrt{r_\mathrm{Y} r_\mathrm{Sb}}/r_\mathrm{X}$. 
Fig.~\ref{fig:tol-facs} shows that indeed, both of these Goldschmidt-like tolerance factors work well for all twenty-five materials in our study with the bounds $0.83 < t_1 <1.28$ and $0.81 < t_2 < 1.25$, respectively.

Finally, we note that, from the above considerations, we expect the above size-based ratios to be useful as well for the more general class of X$_2$YM materials, where M can be any Group V metal or semimetal. Moreover, it is at least plausible that approximately the same numerical bounds for stability will hold when $\rSb$ is replaced with $\rM$.

\subsection{Stable crystal structures} \label{subsec:stable-struct}

With the stability of the X$_2$YSb materials in the $D0_3$ structure understood, 
we next turn to the search for stable structures of the 
materials we find unstable.
In this work, we consider one mono-alkali antimonide and one
bi-alkali antimonide, specifically choosing Cs$_3$Sb and Cs$_2$KSb, which are technologically interesting for photoemission\cite{ref:cultrera_cssb-emittance} and photoabsorption,\cite{ref:ettema-2002, ref:kalarasse_optical} respectively. 
To search for stable structures of these materials, we begin by noting that the most unstable phonon modes occur at the \emph{\textsf{X}}-point of the face-centered-cubic Brillouin zone (see Supplemental Material\cite{ref:supplemental} for phonon dispersion plots), which results in displacements that are commensurate with 
the conventional cubic unit cell (Fig.~\ref{fig:xtal-structs}b). 

Beginning with the simpler material, Cs$_3$Sb, 
we generate 20 different random perturbations of the atoms from the $D0_3$ structure 
in the conventional cubic cell
and then allow the system to relax, as described in Sec.~\ref{sec:procedure}. These 20 perturbed systems all relaxed into just
two distinct structures, only one of which proves to exhibit no imaginary phonon frequencies and thus be mechanically stable. 
Figure~\ref{fig:xtal-structs}c
shows the displacements from the $D0_3$ positions for the final stable structure. The primary displacements occur
for the Cs atoms occupying the Y sites in the X$_2$YSb $D0_3$ structure, which displace along a tetrahedral subset
of the eight possible $\langle$111$\rangle$ directions. We also observe notably smaller displacements on the Sb atoms and on the Cs atoms occupying the X sites. 
Focusing on the largest, Y-site displacements, we note that this displacement pattern can be formed as a linear combination of 
crystal-symmetry respecting 90-degree rotations of the red phonon displacement pattern in Fig.~\ref{fig:xtal-structs}b. 
Thus, the unstable phonon displacements at the \emph{\textsf{X}}-point in the Brillouin zone of the $D0_3$ structure
indeed lead toward the basin of attraction of a stable structure.

Next, we compare the expected peak locations for the powder X-ray diffraction (XRD) pattern from our predicted stable structure with our experimental XRD measurements. (Setup and details described elsewhere.\cite{ref:chess-xrd, ref:Cs-K-Sb-growth}) 
We note that our experimental samples posses texture due to the specific growth method and conditions, which favor certain crystal orientations over others. 
Thus, the relative sizes of the resulting scattering peaks cannot be compared directly to our theoretical calculations, and, moreover, some peaks 
expected from the powder XRD may be missing from our experiments. 
Therefore, the most meaningful comparison to our theoretical results is to compare \emph{only} the peak locations which we \emph{actually observe} in the experiments.
Accordingly, we interpret any measured peaks beyond the calculated $D0_3$ peaks 
to be experimental evidence of breaking of the $D0_3$ symmetry, but we do not necessarily expect to detect all peaks calculated from our predicted structures. 
Finally, to eliminate uncertainty due to subtle differences between density-functional theory and experimental lattice constants, we normalize both our theoretical and experimental peak locations 
so that the 
$hkl=111$ Bragg peak appears exactly at a plane separation of  
$1/\sqrt{3}$ times the cubic lattice constant $a$.

The first two columns of Table~\ref{tab:xrd-peaks} show our measured XRD peaks for Cs$_3$Sb along with the cubic Miller indices $hkl$ corresponding to those peaks.
The last two columns show our theoretical powder XRD peaks for Cs$_3$Sb in the $D0_3$ 
structure and the predicted stable structure, respectively.
As expected, the experimental data does not exhibit clearly all of the theoretical 
peaks. 
The experiment does show an excellent match to the 222 peak expected for both the $D0_3$ and our predicted structure, confirming that the structure in our sample forms a cubic lattice.
Furthermore, we note that the 210 peak does \textit{not} appear in the $D0_3$ structure but \textit{does} appear in our measurements, which clearly indicates symmetry breaking and is consistent with our theoretical findings that this material is not stable in the $D0_3$ structure.
Finally, the predicted stable structure \textit{does} show the 210 peak, consistent with our experiments.

\begin{table}[h!]

\caption{
    Bragg-plane distances $d$ for X-ray scattering peak positions, normalized
    by lattice constant $a$, for cubic Cs$_3$Sb, Cs$_2$KSb, and K$_2$CsSb:
    cubic Miller indices $hkl$ (first column),
    measured $d/a$ for Cs$_3$Sb (second column),
    Cs$_2$KSb (third column),
    and K$_2$CsSb (fourth column),
    calculated $d/a$ for the $D0_3$ structure of Cs$_3$Sb, Cs$_2$KSb, and K$_2$CsSb (fifth column),
    and for the stable structures of Cs$_3$Sb and Cs$_2$KSb (sixth column).
    \label{tab:xrd-peaks}
}

\begin{ruledtabular}

\begin{tabular}{c|c|c|c|c|c}
    $hkl$ & \multicolumn{5}{c}{$d$/$a$}\\
    {}    & \multicolumn{3}{c}{--------- Experiment ---------} & \multicolumn{2}{c}{------------ Theory ------------}\\
    {}    &     Cs$_3$Sb &    Cs$_2$KSb &    K$_2$CsSb &       $D0_3$ & Stable structure \\
    \midrule
      222 &        0.285 &        0.288 &        0.289 &     0.288675 & 0.288675 \\
      311 &           -- &           -- &           -- &     0.301511 & 0.301511 \\
      310 &           -- &           -- &           -- &           -- & 0.316228 \\
      221 &           -- &           -- &           -- &           -- & 0.333333 \\
      220 &           -- &           -- &        0.353 &     0.353553 & 0.353553 \\
      211 &           -- &           -- &           -- &           -- & 0.408248 \\
      210 &        0.436 &        0.453 &           -- &           -- & 0.447214 \\
      200 &           -- &           -- &           -- &     0.500000 & 0.500000 \\
      111 & 1/$\sqrt{3}$ & 1/$\sqrt{3}$ & 1/$\sqrt{3}$ & 1/$\sqrt{3}$ & 1/$\sqrt{3}$
\end{tabular}

\end{ruledtabular}
\end{table}

Having considered the mono-alkali antimonide Cs$_3$Sb, 
we now consider the bi-alkali antimonide Cs$_2$KSb. 
Following the same procedure as for Cs$_3$Sb, we again find a single stable structure with no imaginary phonon frequencies. 
The resulting structure for Cs$_2$KSb 
(Fig.~\ref{fig:xtal-structs}d
shows nearly the same displacement pattern from the $D0_3$ structure as we found for Cs$_3$Sb, with the Y-site atoms (K) assuming the largest displacements (each along one of four tetrahedral $\langle 111\rangle$ directions, 
the pattern which can be anticipated from a displacement pattern [Fig.~\ref{fig:xtal-structs}b, red arrows] from the most unstable phonon mode in the $D0_3$ structure), 
and both the Sb atoms and the X-site atoms (Cs) showing significantly smaller displacements.
Moreover, as can be expected from the above, our theoretical and experimental X-ray diffraction results for this material are indeed very similar to those of Cs$_3$Sb (Table~\ref{tab:xrd-peaks}), so that our predicted stable structure is fully consistent with the structure observed in our experiments, which breaks the $D0_3$ symmetry and exhibits the 210 peak consistent with a cubic lattice.

Finally, as a control case, we have also performed XRD measurements on
K$_2$CsSb, 
a promising electron emitter\cite{ref:bazarov_csksb-emittance, ref:dunham_record-current, ref:schubert_xrd} which we predict to be stable in the $D0_3$ structure. We find that, indeed, the experimental peaks exhibited by this material are different from those found for Cs$_3$Sb and Cs$_2$KSb (Table~\ref{tab:xrd-peaks}) and that they are now fully consistent with expectations for the $D0_3$ structure. 
We thus find full consistency between our experimental measurements and our predictions for both $D0_3$-stable and $D0_3$-unstable materials.

\subsection{Electronic structure and optical absorption properties} \label{subsec:elec-struct}

\begin{figure}[h!]
    \centering
    (a) \\ \includegraphics[width=0.8\linewidth]{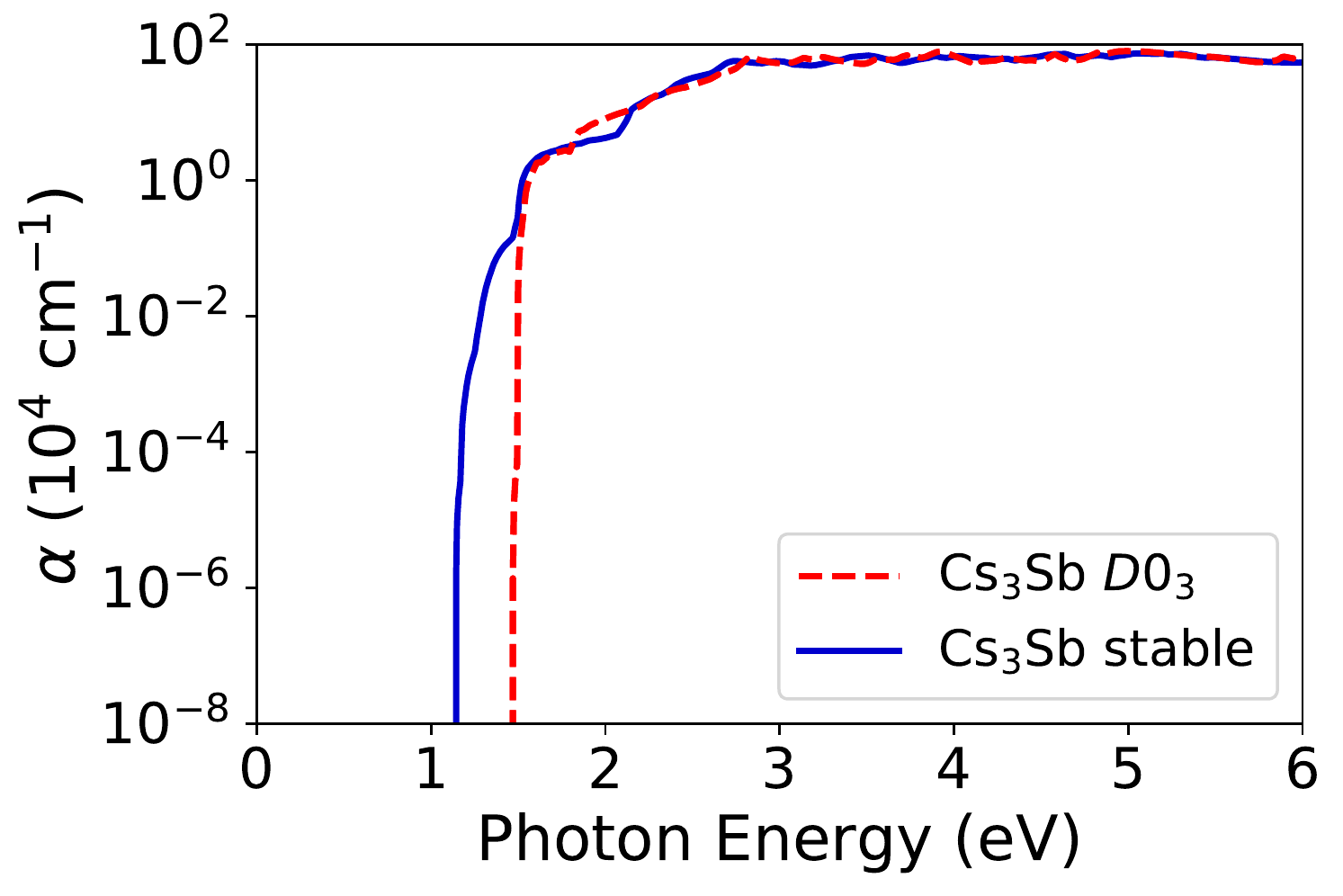}
    \\
    (b) \\ \includegraphics[width=0.8\linewidth]{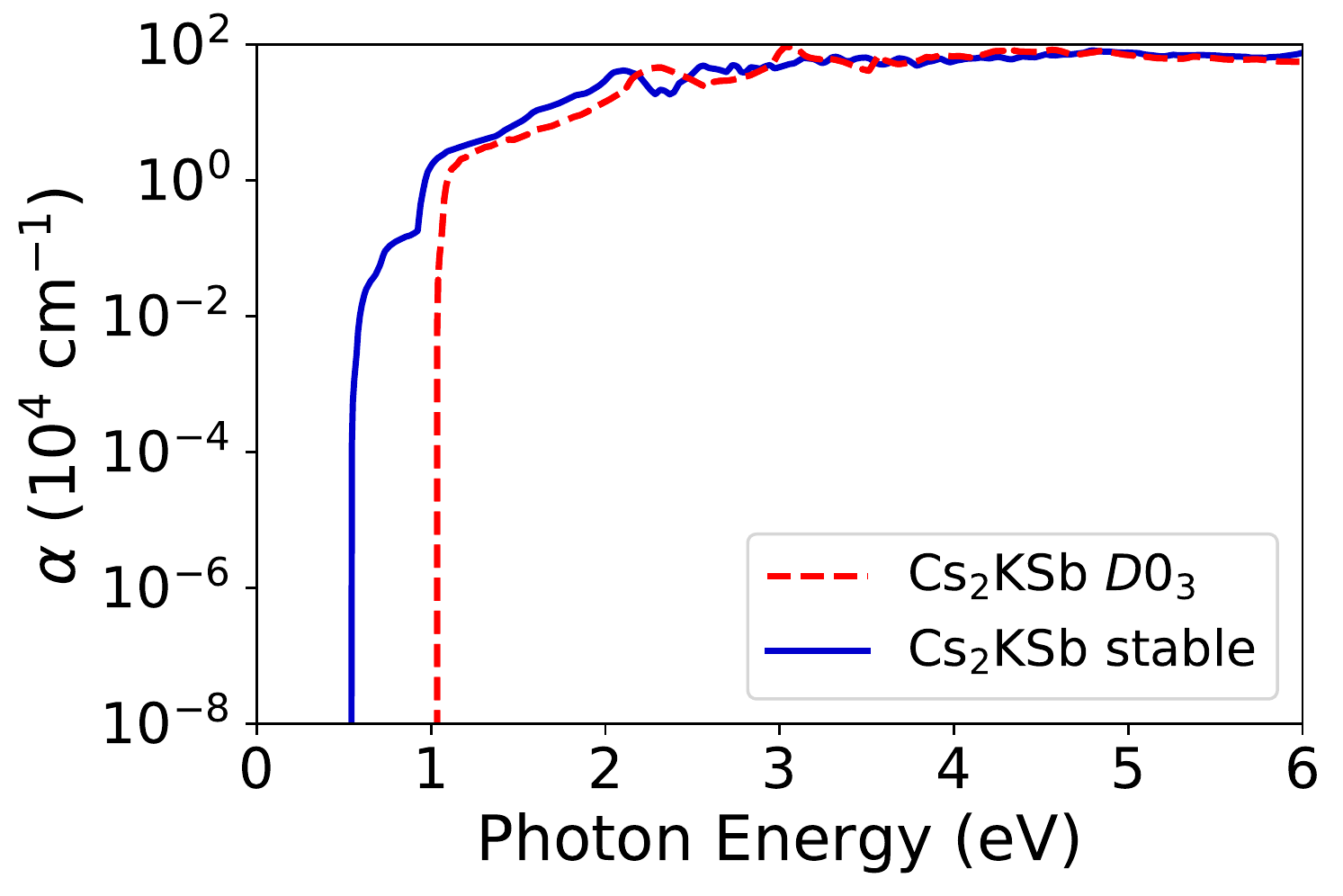}

    \caption{
        Calculated linear absorption coefficient of (a) Cs$_3$Sb and (b) Cs$_2$KSb in
        the $D0_3$ structure (dashed curves) and the stable structures predicted in this
        work (solid curves). 
    }

    \label{fig:alpha}
\end{figure}

Having confirmed experimentally that Cs$_3$Sb and
Cs$_2$KSb indeed break $D0_3$ symmetry, we turn finally to consider the impact of the symmetry breaking on the optical properties of these materials, 
both of which are of interest for photo-applications.\cite{ref:cultrera_cssb-emittance, ref:ettema-2002, ref:kalarasse_optical}  We find that, for both materials, the breaking of symmetry changes the nature of the gap from indirect to direct and lowers the direct optical gap significantly, from  1.53~eV to 1.18~eV and from 1.07~eV to 0.57~eV for Cs$_3$Sb and Cs$_2$KSb, respectively. (See Supplemental Material\cite{ref:supplemental} for more details.)
Finally, Fig.~\ref{fig:alpha} shows our \emph{ab initio} predictions for the optical absorption coefficients for these two materials, where the lowering of the band gaps is evident in the reduction of the location of the absorption edge. 
Above the absorption edge, the absorption of both materials is relatively unaffected by the breaking of the $D0_3$ symmetry due to the fact that the symmetry-breaking displacements are rather small, making the changes in photoexcitation transition rates across the optical gap relatively weak.

\section{Conclusions} \label{sec:conclusion}

This work presents a detailed \textit{ab initio} study of the stability of 25 mono- and bi-alkali antimonides X$_2$YM (X and Y from Group I; M = Sb, a Group V semimetal) in the $D0_3$ structure, the structure commonly assumed in the literature for these materials. 
We find that a significant number of the antimonide materials under study are actually unstable in the $D0_3$ structure and that the instability evident in the imaginary phonon branches can be explained in terms of size mismatches among the atomic radii of the involved species. 
Accordingly, we provide three independent sets of atomic-size criteria that can be used to determine the stability of these compounds: a set of four machine-learning perceptron criteria ($1.00 < \ry$/$\rx < 1.55$ and $0.66< r_\mathrm{M}/\rx< 1.01$), and two different versions of Goldschmidt-like tolerance factors ($0.83<(r_\mathrm{Y} + r_\mathrm{M})/(2r_\mathrm{X})<1.28$ or $0.81<\sqrt{r_\mathrm{Y}r_\mathrm{M}}/r_\mathrm{X}<1.25$).

Finally, for Cs$_3$Sb and Cs$_2$KSb, which prove unstable in the $D0_3$ structure, we identify stable cubic structures, which we find to break $D0_3$ symmetry 
along directions indicated by the unstable phonon displacements at the \emph{\textsf{X}}-point in the face-centered-cubic Brillouin zone. 
These stable structures are consistent with our experimental X-ray scattering data, which indeed indicate broken $D0_3$ symmetry and suggest cubic structure.
Finally, in terms of electronic properties, we note that the nature of the band gap changes from indirect to direct upon breaking of the $D0_3$ symmetry, lowering the optical gaps as reflected in the calculated linear absorption coefficients.

The above results will 
be useful in 
optimizing the growth and ascertaining
the promise of alkali antimonides for next-generation electron 
emitters, for example by clarifying the nature of the primitive unit cell and by allowing more accurate predictions of photoabsorption and photoemission properties.
Future work will include exploration beyond the antimonides (M = Sb) to include
other Group V semimetals and metals (As, Bi).


\begin{acknowledgements}

This work was supported by the U.S. National Science Foundation under Award
    PHY-1549132, the Center for Bright Beams (J.K.N., A.G., J.M.M., S.K.,
    T.A.A.), and by the U.S. Department of Energy, under Contracts No.
    KC0407-ALSJNT-I0013 and No. DE-AC02-98CH1088, and SBIR Grant No.
    DE-SC0009540 and DE-SC0013190 (M.G., J.S.).
The experimental portion of this research used facilities at Cornell High
    Energy Synchrotron Source (CHESS), through support by the NSF and the
    NIH/NIGMS under NSF Grants No. DMR-0936384 and No. DMR-1332208.

\end{acknowledgements}

\FloatBarrier 


\bibliography{refs} 

\end{document}